# About renegades and outgroup-haters: Modelling the link between social influence and intergroup attitudes.

Andreas Flache, University of Groningen, The Netherlands. Department of Sociology / ICS.

Email: a.flache@rug.nl

Polarization between groups is a major topic of contemporary societal debate as well as of research into intergroup relations. Formal modelers of opinion dynamics try to explain how intergroup polarization can arise from simple first principles of interactions within and between groups. Models have been proposed in which intergroup attitudes affect social influence in the form of homophily or xenophobia, elaborated as fixed tendencies of individuals to interact more with in-group members, be more open to influence from in-group members and perhaps even distance oneself from attitudes of outgroup members. While these models can generate polarization between groups, their underlying assumptions curiously neglect a central insight from research on intergroup attitudes. Intergroup attitudes are themselves subject to social influence in interactions with both in- and outgroup members. I extend an existing model of opinion formation with intergroup attitudes, by adding this feedback-effect. I show how this changes model predictions about the process and the conditions of polarization between groups. In particular, it is demonstrated how the model implies that intergroup polarization can become less likely if intergroup attitudes change under social influence; and how more complex patterns of intergroup relations emerge. Especially, a renegade minority ('outgroup-lovers') can have a key role in avoiding mutually negative intergroup relations and even elicit 'attitude reversal', resulting in a majority of individuals developing a negative attitude towards their in-group and a positive one of the outgroup. Interpretations of these theoretical results and directions for future research are further discussed.

**About renegades and outgroup-haters: Modelling the link between social influence and intergroup attitudes.**

1. Introduction

It is a central question in research on opinion dynamics whether and under what conditions societies polarize, falling apart into a small number of deeply antagonistic factions. Recent developments that underscore the importance of this question are, for example, the rise of right-wing populist parties in many Western societies, the Brexit referendum in the U.K., or the trend towards deepening divisions in the political landscape of the U.S. (Abramowitz and Saunders 2008; DiMaggio, Evans, and Bryson 1996; Evans 2003; Fiorina and Abrams 2008; Gentzkow 2016; Liu and Srivastava 2015).

Research on intergroup relations points to negative intergroup attitudes as one of the important sources of opinion divisions in society. Intergroup attitudes are positive or negative attitudes individuals hold about different groups in their society, including their own ingroup. Negative attitudes towards outgroups can for example arise from feelings of threat individuals perceive in the face of increasing immigration or rising ethnic diversity in their social environment (McLaren 2003; Moody 2001; Quillian 1995; Semyonov, Raijman, and Gorodzeisky 2006; Vervoort, Scholte, and Scheepers 2011). Negative outgroup attitudes can also form when people generalize from experiences with individual members of outgroups toward their attitude about the outgroup as whole. Studies in the tradition of contact theory (Allport 1954; Pettigrew and Tropp 2006, 2013), as well as research on attitude generalization and stereotype change (Johnston and Hewstone 1992; Kunda and Oleson 1997; Paolini, Crisp, and McIntyre 2009) demonstrated this generalization for positive interpersonal experiences as well as negative ones (Dolderer, Mummendey, and Rothermund 2009; Stark, Flache, and Veenstra 2013).

Attitudes about a group not only arise from interpersonal interactions and perceptions of threat, but they also shape how individuals interact with and are influenced by members of different groups. This role of intergroup attitudes has been recognized by researchers developing formal models of the dynamics of opinion formation. A number of models, e.g. (Baldassarri and Bearman 2007; Flache and Macy 2011; Jager and Amblard 2005; Macy et al. 2003; Mark 2003) incorporated the possibility of repulsive social influence, assuming that individuals strive to be dissimilar to people they dislike, and accentuate disagreement with others if these are perceived as being too discrepant (also called 'xenophobia').  Building on research on intergroup dynamics and social identity (Brewer 1991; Tajfel 1978), further models (Salzarulo 2006) assume that individuals may change their opinion to adopt a position prototypical for their in-group or to distance themselves from an opinion perceived as prototypical for an outgroup (see also (Huet and Deffuant 2010; Huet, Deffuant, and Jager 2008) for a similar approach).  It has in particular been assumed that perceived dissimilarity not only arises from disagreement in opinions between individuals, but also from 'demographic' differences between individuals representing fixed characteristics like gender, ethnicity, or race (Feliciani, Flache, and Tolsma 2017; Flache and Mäs 2008a, 2008b; Grow and Flache 2011; Macy et al. 2003; Mäs et al. 2013). These models show how demographic diversity can give rise to polarization between demographic groups, in particular when these groups differ in a number of demographic dimensions at the same time, generating "demographic faultlines" (Lau and Murnighan 1998) in a population.

Formal models of opinion dynamics take into account that negative intergroup attitudes can foster disagreement and polarization between groups, but to my knowledge existing models curiously neglect that intergroup attitudes are at the same time themselves subject to social influence and are shaped by opinion dynamics. Empirical research on effects of direct and extended intergroup contact (Munniksma et al. 2013; Pettigrew and Tropp 2006) has established how attitudes about groups are shaped by interactions both between members of the same group and members of different groups, via direct or indirect experiences with group members, or via social influence (Van Zalk et al. 2013).

If we consider in models of opinion dynamics that intergroup attitudes not only affect social influence between individuals but are simultaneously shaped by social influence processes, different theoretical implications can arise about the role that intergroup attitudes have for societal polarization. For example, if xenophobia and its counterpart of ingroup-favoritism (Sherif 1966) are a 'hardwired' element in social influence, models have an inherent tendency to generate opinion polarization that divides a population along the lines of its salient demographic divisions. However, if xenophobia can be 'unlearned', for example when individuals in different groups discover sufficient ground for agreement and can form mutually positive relations, consensus in a society could be more robust against increasing demographic diversity than suggested by models with hardwired xenophobia. Also, as research on intergroup attitudes has pointed out, people can sometimes have negative attitudes towards their own group and prefer outgroups above their ingroup, for example if their ingroup has low social status in their society, or is perceived to violate important personal norms (Becker and Tausch 2014).

The presence of such 'renegades' could profoundly change polarization dynamics when intergroup attitudes are open to influence. Renegades can influence their fellow ingroup-members to adopt a more critical perspective on the own group and think more positively about outgroups. This might prevent intergroup polarization where it would develop otherwise if intergroup attitudes were fixed and xenophobic. But the opposite is also possible. A small critical mass of 'outgroup-haters' might convince a moderate majority within their own group to become more critical of the outgroup, fuelling in the process increasing disagreement between groups that eventually gives rise to polarization between groups.

To explore these possibilities, the current paper proposes a simple model that extends previous models of social influence dynamics which assumed fixed outgroup attitudes. I integrate the assumption that intergroup attitudes are subject to social influence and analyze with computational experiments how conditions and dynamics of opinion polarization change if intergroup attitudes are assumed to be flexible and socially influenced, rather than static and hardwired. A description of the model will be given in Section 2. Section 3 presents the results of several simulation experiments. The paper concludes with a discussion of the substantive interpretation of results and possible directions for future research in Section 4.

## 2. Model

The model presented here draws upon and extends earlier models that combine assimilative and repulsive social influence in opinion. To concentrate on the effects of introducing socially influenced intergroup attitudes, other aspects of the model are chosen to be maximally simple. To begin with, the population of *N* individuals consists of only two distinct groups and group membership is stable over time. Every individual *i* is member of either group 0 or group 1, indicated by its group membership $g_i \in \{0,1\}$. Furthermore, individuals influence each other with regard to only *one* opinion dimension *o*. The key difference to earlier models is that this opinion represents the intergroup attitude, which in turn affects how individuals are influenced by in-group members and outgroup members. More precisely, the opinion value $o_{it}$ expresses the attitude an individual *i* has at time point *t* about group 1 relative to group 0, where higher values indicate a more positive attitude towards group 1 and a less positive attitude towards group 0. Following many earlier models, I constrain the value range of opinions, assuming $0 \leq o_{it} \leq 1$. If $o_{it} < 0.5$, individual *i* has a more positive attitude towards group 0 than towards group 1, while $o_{it} > 0.5$ indicates the opposite relation. A value of $o_{it} = 0.5$ expresses indifference between the two groups. It should be noted that for members of group 1, the opinion *o* models their evaluation of the in-group relative to the outgroup, whereas for members of group 0, it captures the evaluation of the outgroup relative to the in-group.

Dynamics of the model unfold in a sequence of consecutive interaction events, in which in every event a pair of two different population members *i* and *j* is selected with equal probability from all possible pairs in the population for an interaction. Notice that this imposes the simplification that interactions are not constrained by network structure or spatial distances. The 'interaction regime' here draws upon the pairwise interaction mechanism introduced by Deffuant and co-authors (Deffuant et al. 2000).

All individuals *k* who are not involved in an interaction at time point *t* do not change their opinions, thus $o_{k,t+1} = o_{k,t}$. If *i* and *j* do interact, then both can modify their current opinions to move closer towards or away from the opinion of the interaction partner as given by equations 1 and 2.

$$o_{i,t+1} = o_{it} + \Delta o_{it} = o_{it} + \mu w_{ijt} (o_{jt} - o_{it}) \qquad [1]$$

$$o_{j,t+1} = o_{jt} + \Delta o_{jt} = o_{jt} + \mu w_{jit} (o_{it} - o_{jt}) \qquad [2]$$

The parameter $\mu$ ($0 < \mu \leq 0.5$) in [1] and [2] defines the rate of opinion change. The influence weights $w_{ijt}$ and $w_{jit}$ represent the direction and magnitude of the influence of *i* on *j* and *j* on *i*, respectively. Weights are constrained by $-1 \leq w_{ij} \leq 1$. A positive weight $w_{k,m}$ entails assimilative influence (*k* moving her opinion closer towards *m*'s opinion), whereas a negative weight imposes differentiation (*k* moving her opinion away from *m*'s opinion). With $w_{k,m} = 0$, *k* does not change her opinion, reflecting indifference towards the source of influence, *m*. In this basic form, equations 1 and 2 allow interactions to push the opinion outside of the opinion interval [0,1] if weights are negative. In this case, the resulting opinion is truncated to the interval boundary that was crossed by the opinion shift. In some models that combine assimilation and differentiation, opinions are constrained with more sophisticated approaches (e.g. Flache and Macy 2011; Flache and Mäs 2008a,b; Huet and Deffuant 2010), but this seems to have little effect on the main model dynamics (Feliciani et al. 2017).

The link between intergroup attitudes and social influence is implemented in the way how weights are computed. I adopt a two-step process. In the first step, the discrepancy $d_{ijt}$ is computed that individual *i* experiences at time point *t* between herself and individual *j*. Discrepancy is based on the current level of disagreement $|o_{jt} - o_{it}|$, whether *i* and *j* belong to the same group ($|e_j - e_i|$) and on the attitude that *i* currently has towards the group to which *j* belongs. Discrepancy is constrained by $0 \leq d_{ijt} \leq 1$. In the second step, the actual influence weight $w_{ijt}$ is obtained from a weight function that modifies the discrepancy $d_{ijt}$. This function is monotonous but not necessarily linear. Broadly, the larger the discrepancy, the lower the influence weight.

More precisely, discrepancy $d_{ijt}$ is computed as given by equation 3.

$$d_{ijt} = \beta_O |o_{jt} - o_{it}| + \beta_D |e_j - e_i| + \beta_A [e_j(1 - o_{it}) + (1 - e_j)o_{it}] \quad [3]$$

The parameters $\beta_O, \beta_D, \beta_A$ in [3] scale the relative impact that respectively, opinion disagreement, demographic differences and the intergroup attitude towards *j*'s group have on $d_{ijt}$. I impose the constraint $\beta_O + \beta_D + \beta_A = 1$. If intergroup attitudes play no role ($\beta_A = 0$), this implementation is compatible with earlier models that incorporate only opinion disagreement and fixed xenophobia in otherwise the same framework (e.g. Feliciani et al. 2017; Flache and Mäs 2008a,b). The new element of intergroup attitudes is added such that a more positive attitude $o_{it}$ of *i* towards group 1 means less discrepancy if the interaction partner *j* belongs to group 1, whereas it means more discrepancy if *j* belongs to group 0. Conversely, if *j* belongs to group 0, then a higher value of $o_{it}$ implies more discrepancy and a lower value means less discrepancy.

The actual influence weight $w_{ijt}$ is computed from the discrepancy $d_{ijt}$ with a non-linear weight function *f* adapted from (Mäs, Flache, and Kitts 2014), as given by equation [4] and visualized by Figure 1. As figure 1 shows, this weight function implements a non-linear relation between discrepancy and the impact that a source of influence *j* has on *i*'s opinion. Similar to earlier implementations (e.g. Jager and Amblard 2005) it is assumed that if the level of discrepancy is such that the opposing psychological motivations towards assimilation and differentiation are roughly in balance, changes in discrepancy have relatively little effect. More technically, the function *f* yields a weight that is positive if discrepancy falls below a threshold level of 0.5, the midpoint of the range of possible discrepancy levels. The weight turns negative once discrepancy exceeds this threshold. Changes in discrepancy close to the threshold level have only a relatively small impact on the weight such that the influence weight is close to zero in this region. The weight changes more steeply in discrepancy as *d* approaches the boundaries of its theoretically possible range of [0,1].

$$f(d) = \begin{cases} (1 - 2|d|)^2, & d < \frac{1}{2} \\ -(2|d| - 1)^2, & d \geq \frac{1}{2} \end{cases} \quad [4]$$

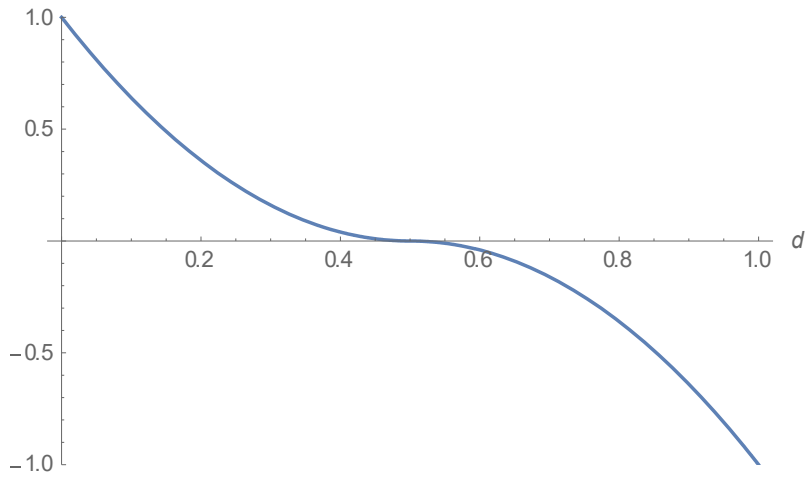

**Figure 1. Influence weight *w* as function of discrepancy *d* (equation 4)**

### 3. Results

I present here results from a set of computational experiments[1] that explore whether and if so how conditions and dynamics of polarization change, if we model intergroup attitudes as socially influenced rather than given by 'fixed xenophobia'. More precisely, opinion dynamics will be compared across different combinations of the parameters $\beta_O, \beta_D, \beta_A$. The results presented in this paper are the first step of a fuller exploration of the behavior of the model here proposed. Thus, I keep a number of conditions constant across all the experiments reported below. To begin with, population size is assumed to be constant throughout and fixed at $N=100$, where groups 0 and 1 have equal size ($N_0=N_1=50$). The rate of opinion change is fixed at $\mu=0.5$. Further constant conditions were discussed above and include that there is only one opinion dimension *o*, and that interaction is equally likely for every pair of population members.

The key outcome of interest in the simulation experiments is the degree of polarization both within and between the two groups. To assess between-group polarization, I measure the signed difference between the mean opinions in both groups, $|\overline{o_{g=0}} - \overline{o_{g=1}}|$. If the absolute difference in group means is close to one, this is a clear sign of strong between-group polarization. A low difference between the mean opinions of the groups, however, does not necessarily show that there is no polarization in the population as a whole. The population can also fall apart into two opposed factions containing members of both groups. To distinguish this form of 'population polarization' from between-group polarization, I use a measure of population polarization adapted from (A Flache and Macy 2011). Population polarization $P_t$ at time point *t* is derived from the variance of all pairwise opinion distances in the population as given by equation 5.

$$P_t = \frac{4}{N^2} \sum_{ij}^{i=N, j=N} (|o_{it} - o_{jt}| - \overline{d_t})^2, \qquad [5]$$

---

[1] The simulations, all visualizations and analyses of results in this paper were implemented and conducted with Wolfram Mathematica© Version 11.0.1.0.

In equation 5, $\overline{d_t}$ denotes the average opinion distance across all pairs (*ij*) in the population. The minimum level of polarization (*P* = 0) obtains when all pairwise distances are zero, corresponding to full consensus in the population[2]. *P* obtains its maximal value of 1 if the population is split into two equally large factions with maximal mutual disagreement and full agreement within each of the factions. It is also useful to assess to what extent polarization occurs within each of the two groups separately. For this, the measure $P_t$ is computed separately for each of the two groups ($P_{t,g=0}$, $P_{t,g=1}$).

**Simulation experiment 1: the relative weight of intergroup attitudes in social influence**

Previous modelling work has shown how fixed xenophobia can increase opinion polarization between groups, particularly when the demographic differences are clear and salient (e.g. Feliciani et al. 2017; Flache and Mäs 2008a,b). This suggests that the stronger the relative weight $\beta_D$ for discrepancy between individuals, the more likely attitudes in the population polarize between the two subgroups. In experiment 1, it will be assessed whether this result changes if intergroup attitudes are themselves subject to social influence. I start from an initial opinion distribution that imposes a mild ingroup-bias in both groups. This reflects the notion of ingroup favoritism. Technically, the initial opinions in both groups are randomly drawn from two Beta distributions that are symmetric around the midpoint of the opinion interval, with Beta(7.5,10) for group 0 and Beta(10,7.5) for group 1. The mean opinion in this initial distribution is about 0.43 for group 0 and about 0.57 for group 1. Initial opinions in both groups have the same expected standard deviation of approximately 0.12. Figure 2 visualizes the corresponding probability density functions together with the aggregate distribution for the population as a whole.

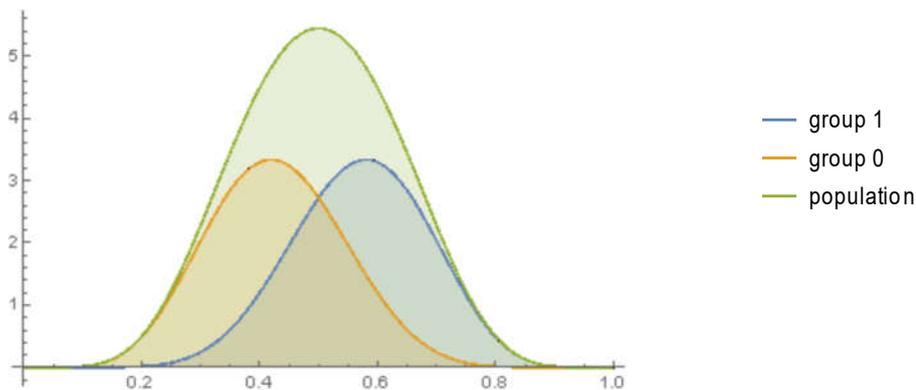

**Figure 2. Initial opinion distribution in experiment 1 (group 1: Beta(10,7.5), group 0: Beta(7.5,10)).**

In experiment 1, the relative impact of the intergroup attitude on discrepancy is systematically varied across the entire range of possible values. Two conditions are distinguished. First, in the 'fixed xenophobia' condition, the parameter that is manipulated is $\beta_D$, keeping $\beta_A = 0$ and $\beta_O + \beta_D = 1$. In the 'socially influenced intergroup attitude' condition, the parameter manipulated is $\beta_A$, keeping $\beta_D = 0$ and $\beta_O + \beta_A = 1$. In both conditions, the manipulated parameter is varied from 0 to 1 in steps of 0.025, yielding 41 different levels of impact of intergroup attitudes. With $\beta_D$=0 or $\beta_A = 0$, discrepancy between two individuals *i* and *j* depends exclusively on their current level of disagreement in their attitude towards group 1, $|o_{jt} - o_{it}|$. At the other end of the spectrum, with $\beta_D=1$ or $\beta_A = 1$, disagreement is irrelevant for the discrepancy between *i* and *j*. Here, only demographic difference (fixed xenophobia condition) or *i*'s attitude towards *j*'s group (socially influenced intergroup attitude condition) matter for the direction and magnitude of influence of *j* on *i*.

---

[2] Other than in Flache and Macy (2011), the set of pairs includes here also the self-distances in the pairs *(i,i)*. While these are known to be zero by definition, including self-distances simplifies definition and interpretation of the measure.

Figure 3 shows the effect of $\beta_D$ on indicators of between-group polarization in the fixed xenophobia condition. The results shown are averages of the outcomes of 100 independent realizations of the simulation per level of $\beta_D$, measured after 30.000 simulation events.

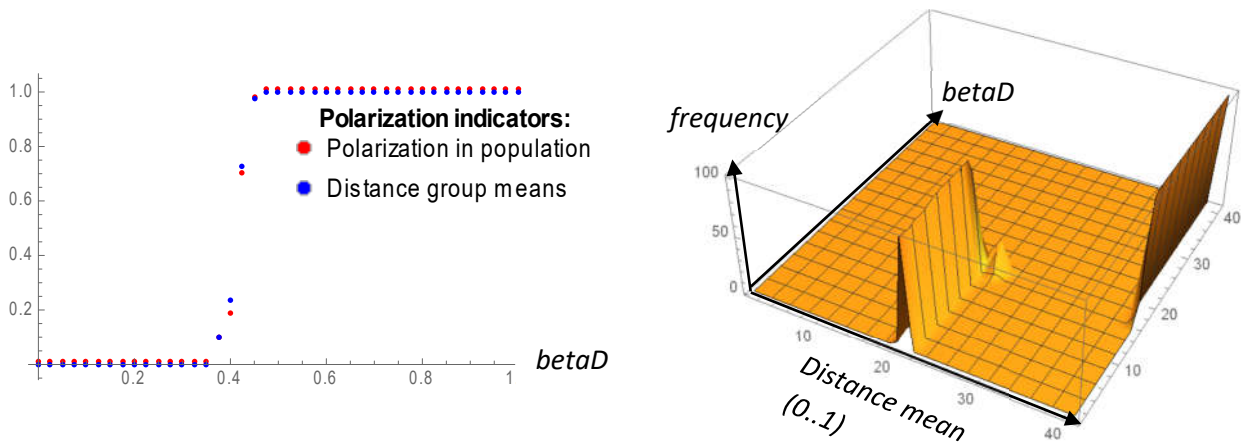

**Figure 3. Effects of $\beta_D$ on between-group polarization in fixed xenophobia condition of experiment 1. Left: Polarization indicators averaged across 100 realizations per level of $\beta_D$. Right: distribution of distances between mean opinions per group. Results after 30.000 simulation events per run.**

Figure 3 shows how the population shifts from a consensus regime into a regime with maximal between-group polarization when fixed xenophobia exceeds a critical threshold level of approximately $\beta_D$=0.4. The left graph shows a sharp increase in the average measures of the mean opinion distance between the groups, $|\overline{o_{g=0}} - \overline{o_{g=1}}|$, and of the level of polarization in the population, $P_t$, when $\beta_D$ passes the critical level. Below that point, both outcome indicators yield zero, indicating that after 30.000 iterations all population members have adopted the same intergroup attitudes and there are no discernable differences between the two groups. Despite an initial distance between the group means of about 0.145, in this region of the parameter space the initial disagreement within and between groups is not sufficiently large to trigger repulsive influence dynamics that could counter the pressures towards assimilation. Eventually all individuals converge to an opinion close to the initial population mean value of 0.5. However, when $\beta_D$ exceeds the critical level, only little disagreement is needed to give individuals a negative influence weight *w* towards each other if they do not belong to the same group. For example, for $\beta_D = 0.4$, negative influence is triggered between individuals from different groups if their disagreement $|o_{jt} - o_{it}|$ exceeds a level of approximately 0.15, whereas for members of the same group only a disagreement of 0.8 or more can elicit repulsive influence. In this region of the parameter space, the initial between-group differences are large enough to trigger repulsive influence dynamics between members of different groups, pushing their attitudes increasingly away from those of the outgroup towards the extreme poles of *o*=0 and *o*=1 for group 0 and 1, respectively. Assimilation within the groups assures that all group members eventually agree with these extremely negative outgroup attitudes. The close match of the two polarization indicators in the left part of figure 3 further confirms that all polarization is between the groups and not within groups ($P_{t,g=0}$, $P_{t,g=1}$ are approximately zero after 30.000 simulation events).

Figure 4 shows results for the condition in which intergroup attitudes are socially influenced. It visualizes the effect of $\beta_A$ on average indicators of between-group polarization and their distribution across independent simulation runs (100 independent realizations, results measured after 30.000 simulation events).

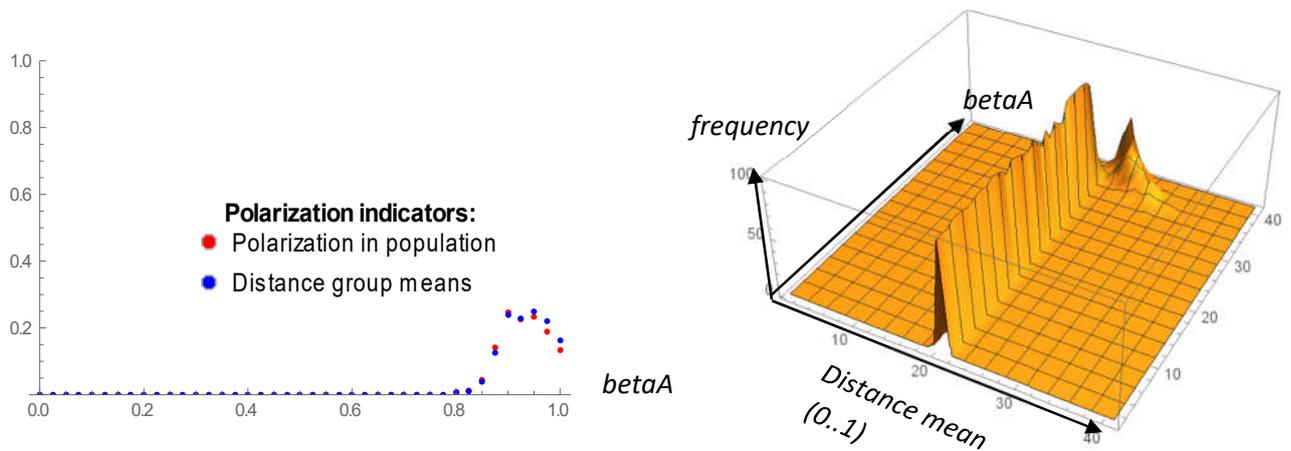

**Figure 4. Effects of $\beta_A$ on between-group polarization in condition with socially influenced intergroup attitudes of experiment 1. Left: Polarization indicators averaged across 100 realizations per level of $\beta_A$. Right: distribution of distances between mean opinions per group. Results after 30.000 simulation events per run.**

A comparison of figures 3 and 4 reveals clear differences. One commonality between both conditions of experiment 1 is that opinion dynamics result in consensus if the impact of intergroup attitudes falls below a critical level. However, for $\beta_A$ a level of at least 0.8 must be reached before the population leaves the consensus regime. Moreover, the degree of intergroup polarization is much lower than for fixed xenophobia, even when $\beta_A$ approaches its maximum level of 1.0. The right part of figure 4 shows that also in this region none of the runs with socially influenced intergroup attitudes has generated perfect between-group polarization. Finally, we observe that the effect of $\beta_A$ on polarization is not monotonous. The left part of figure 4 shows how both polarization indicators peak at approximately $\beta_A$=0.9 at a level around 0.25, and then decline down to about 0.2 when $\beta_A$ further increases.

Why is the effect of $\beta_A$ on between-group polarization so different from the effect of $\beta_D$? Below the critical level of approximately $\beta_A$=0.8, opinion dynamics are dominated by assimilation. As we see, much higher levels of $\beta_A$ than of $\beta_D$ are required to leave the consensus regime. This follows directly from the discrepancy function [3]. If $\beta_A=\beta_D$, then the 'xenophobia term' $\beta_D|e_j - e_i|$ always adds more to the discrepancy experienced in an interaction with an outgroup member than the 'attitude term' $\beta_A[e_j(1 - o_{it}) + (1 - e_j)o_{it}]$ does, unless *i*'s attitude towards the outgroup is maximally negative. Given that intergroup attitudes are only mildly biased at the outset, this explains the narrower range of levels of $\beta_A$ in which the population escapes from the pull of consensus. But this does not yet explain the non-monotonous effect that we observe for $\beta_A$. To shed more light on the different dynamics of intergroup polarization with socially influenced intergroup attitudes, figure 5 visualizes the change of the opinion distributions over time for two typical runs with $\beta_A = 0.9$ and $\beta_A = 1$, respectively.

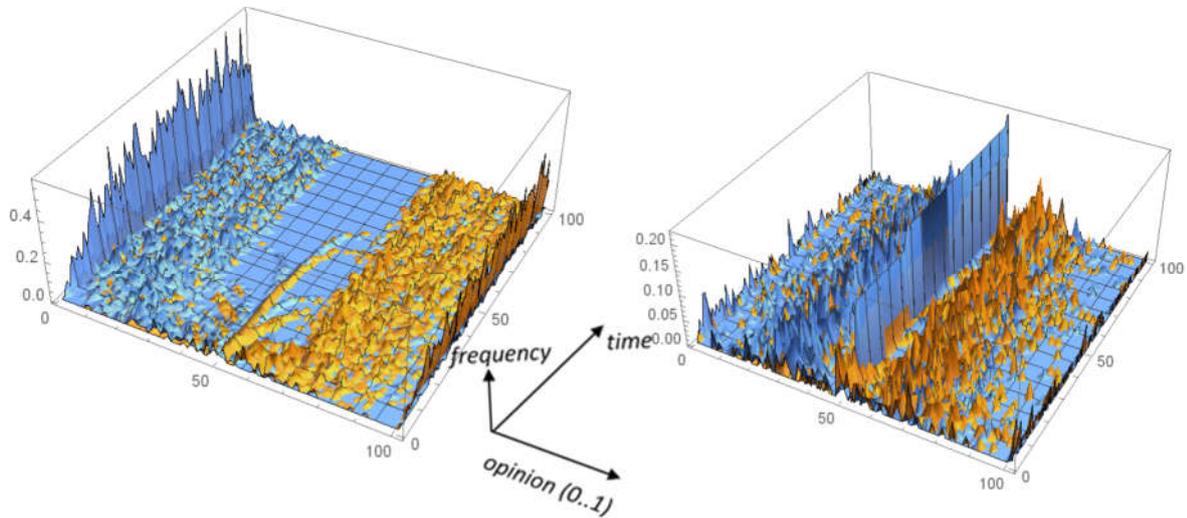

**Figure 5. Change of opinion distribution in 2 typical runs in experiment 1, for $\boldsymbol{\beta_A = 0.9}$ (left) and $\boldsymbol{\beta_A = 1}$ (right) in the condition with socially influenced intergroup attitudes ($\boldsymbol{\beta_D = 0}$). Blue: group 0, Orange: group 1.**

Figure 5 shows how between group polarization in this condition is relatively low compared to fixed xenophobia and how it is lower for $\beta_A = 1$ than for $\beta_A = 0.9$. In both runs, the frequency distribution of opinions shows that there is no clear separation between members of the two groups. In the run for $\beta_A = 0.9$ we find two peaks at the extreme poles of 0 and 1, but unlike for fixed xenophobia, we see a pattern of 'mixed polarization' in which both groups contain extreme outgroup-haters as well as 'renegades' who dislike their own group. Correspondingly, the difference in mean values between the two groups is rather low even after 30.000 iterations. With $\beta_A = 1$, there is even less of a tendency towards polarization between the groups. Instead, the opinion distribution shows after some time a clear peak at the moderate attitude of *o*=0.5 which is adopted by members of both groups. Finally, both runs reveal that around the peaks of the distribution there is a fair amount of random movement so that dynamics have not settled down to an equilibrium state within the time frame simulated here.

Figure 6 further describes the overall tendencies in these dynamics. The figure presents the average frequencies of the opinion distributions that evolved across 100 independent realizations in the two conditions $\beta_A = 0.9$ and $\beta_A = 1$.

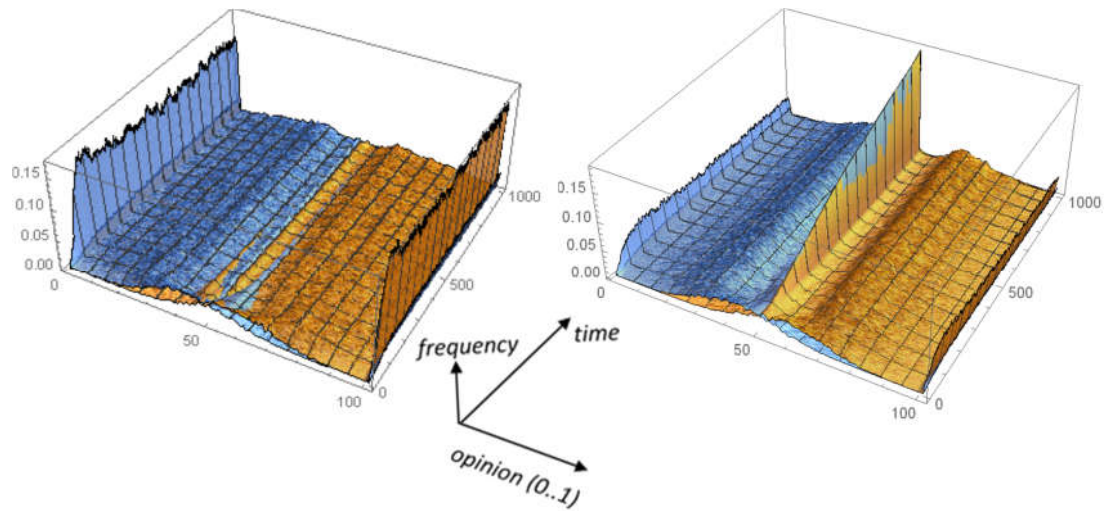

**Figure 6. Change of aggregated opinion distributions over 100 independent realizations for $\beta_A = 0.9$ (left) and $\beta_A = 1$ (right) in the condition with socially influenced intergroup attitudes ($\beta_D = 0$). Blue: group 0, Orange: group 1.**

Figure 6 shows that there is some extent of polarization between groups both for $\beta_A = 0.9$ and for $\beta_A = 1$, in the sense that in both conditions the emergent aggregated opinion distributions show peaks at both extreme poles. While these extremal peaks are much more pronounced for $\beta_A = 0.9$, the peak around the neutral attitude 0.5 is clearly more pronounced for $\beta_A = 1$. At the same time, in both conditions members of both groups adopt positions across the entire spectrum, including extreme ingroup-lovers, ingroup-haters and moderates. Especially for $\beta_A = 1$ smaller peaks arise within both groups at attitudes that are moderately in favor of the ingroup. Inspection of the dynamics suggests that for individuals at these positions, different opposing pressures to modify their attitudes tend to cancel each other out, such that they can retain their moderate pro-ingroup position at least temporarily. The pressures balancing each other are the opposing attractions towards maximally extreme ingroup-lovers and moderates within their own group, and the push towards differentiation from moderate and extreme outgroup members. Overall, in line with the moderate ingroup-bias that was imposed in the initial distribution, members of both groups remain on average slightly more positive about their ingroup than about the outgroup and occur more frequently at the pole of extreme ingroup-love than at the pole of maximal ingroup-hate.

Inspection of the trajectories of single individuals in single simulation runs gives more insight into why some individuals turn into ingroup lovers and others into ingroup haters. Figure 7 below shows the evolution of the opinions of all individuals in the population in a specific run. The left part of the figure shows how in both groups individuals who initially have a more positive ingroup attitude (*o* > 0.5 for group 1, *o* < 0.5 for group 0) are likely to move towards the pole of a maximally positive ingroup attitude, while those who initially are more critical (*o* < 0.5 for group 1, *o* > 0.5 for group 0) tend to move in the opposite direction.

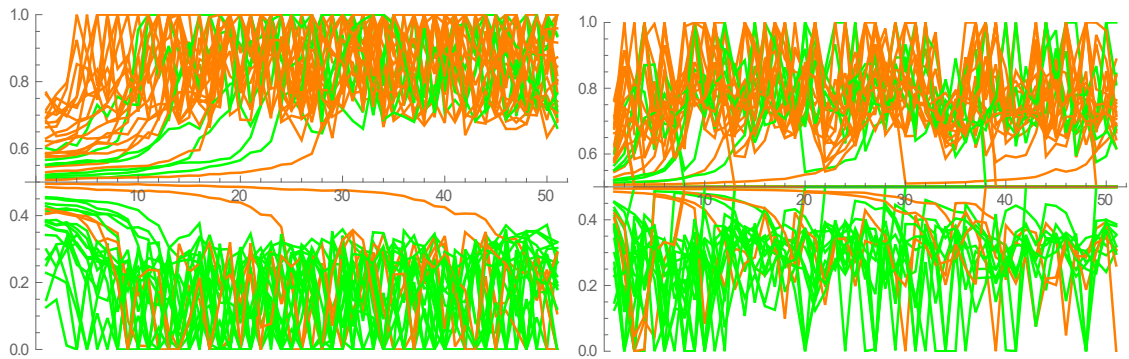

**Figure 7. Opinion trajectories over time for typical runs for $\beta_A = 0.9$ (left) and $\beta_A = 1$ (right) in the condition with socially influenced intergroup attitudes ($\beta_D = 0$). Green: group 0, Orange: group 1.**

With $\beta_A$ close to or equal to 1.0, the opinion movement of a particular individual is primarily dominated by her intergroup attitude. Individuals who have a more favorable attitude of their ingroup (*o* > 0.5 for group 1, *o* < 0.5 for group 0) are moving towards the opinion of an ingroup member if they interact with one, and move away from the opinion of outgroup members they encounter. Given that in the initial distribution the average attitude of group 1 is above 0.5 while the average attitude of group 0 is below 0.5, this explains why most group 1 members move upwards and most group 0 members move downwards in the dynamics shown by figure 7. As these individuals feel even more positively about their ingroup in the process, their initial tendency becomes self-reinforcing, pushing them towards the boundaries of maximal ingroup-love combined with maximal outgroup-hate. Conversely, the minority of initial 'renegades' in both groups (*o* < 0.5 for group 1, *o* > 0.5 for group 0) is on average attracted towards the mean opinion of the outgroup and shifts away from the attitudes of their ingroup members. As a consequence, these individuals become increasingly critical of their ingroup and tend to move towards the opposite extreme.

But why do the dynamics not settle into an equilibrium state within the time frame we observe? The reason is that individuals who strongly like their ingroup and thus dislike the outgroup, are not happy with having the same attitude than 'renegade' outgroup members have who agree with them. At the same time, these renegade outgroup members 'chase' them in the attitude space because they like to agree with members of the outgroup they love. Once an ingroup-lover of one group interacts with an ingroup-hater of the other group, the ingroup lover has a negative influence weight towards the member of the other group and will therefore - paradoxically - be pushed to shift away from his prior positive attitude towards his own group. However, given that ingroup-lovers occur more frequently on both sides than ingroup-haters do, it is likely that the ingroup-lover will be pulled back sooner or later after an interaction with an ingroup-member. Figure 7 shows that these destabilizing dynamics occur for both $\beta_A = 0.9$ and $\beta_A = 1$, but on average many more individuals adopt a moderate attitude at any point in time for $\beta_A = 1$ and many more extremist attitudes occur for $\beta_A = 0.9$. How this difference comes about can be further seen from inspecting the influence weight functions towards in- and outgroup members in both conditions. Figure 8 depicts the influence weight as a function of the intergroup attitude and of the opinion disagreement between two individuals who interact, separately for $\beta_A = 0.9$ and $\beta_A = 1$.

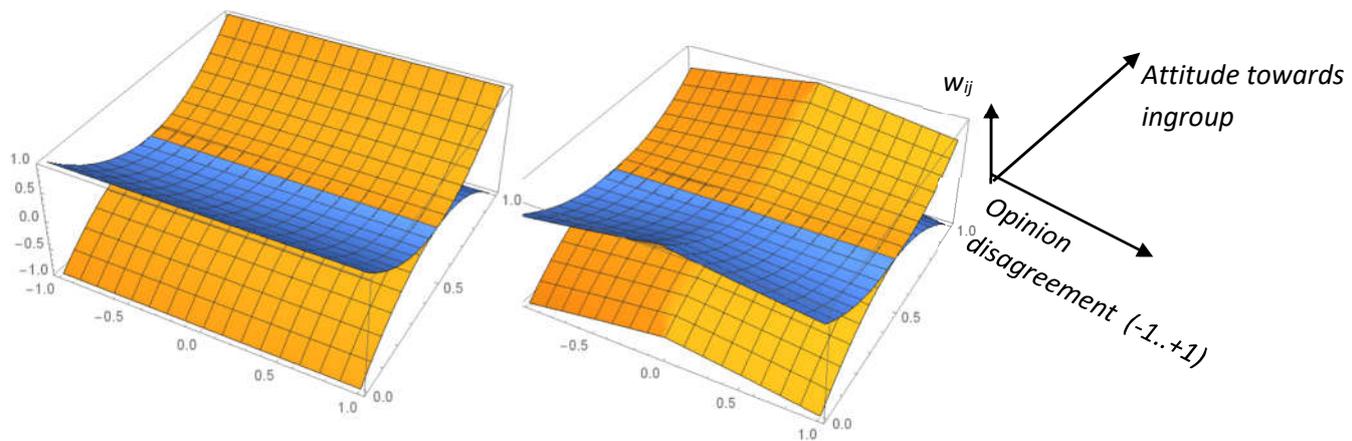

**Figure 8. Influence weights towards ingroup and outgroup member for $\beta_A = 0.9$ (right) and $\beta_A = 1$ (left) in the condition with socially influenced intergroup attitudes ($\beta_D = 0$). Orange: ingroup weight, Blue: outgroup weight.**

Figure 8 explains why there is such a strong tendency for individuals to adopt a moderate intergroup attitude if $\beta_A = 1$. If $\beta_A = 1$ and actor *i* has a neutral attitude (*o*=0.5) then the influence weight *w*<sub>ij</sub> is zero towards both ingroup- and outgroup members, regardless of the level of disagreement between *i* and *j*. Thus, once random movements have led an individual into the position *o*=0.5, this individual will no longer change her attitude. This is different if $\beta_A = 0.9$. Here, disagreement still affects the influence weight. Although with a moderate intergroup attitude of *o*=0.5, individuals have the same influence weight towards members of both groups, a neutral influence weight of *w*=0 can only obtain if the level of disagreement between *i* and *j* is exactly 0.5 for both ingroup and outgroup-members *j*. It is impossible to obtain a configuration in which more than one individual in the population can at the same time disagree by 0.5 with both all ingroup and all outgroup-members, so that for $\beta_A = 0.9$ we do not observe the emergence of a stable attitude of *o*=0.5. By contrast, *o*=0.5 is always an equilibrium attitude for *i* if $\beta_A = 1$. An analytical proof for this can be found in the appendix.

To sum up, experiment 1 has shown distinctive differences between fixed xenophobia and socially influenced intergroup attitudes. Overall, the model with socially influenced intergroup attitudes generates between group polarization in a much narrower range of conditions. For those conditions where some intergroup polarization emerges, it does so to a much smaller extent than under the model with fixed xenophobia. Moreover, the dynamics and opinion distributions that emerge if there is no consensus (approximately for $\beta_A > 0.8$ ), are more complex and richer than those we find for higher levels of fixed xenophobia. With fixed xenophobia above a level of approximately $\beta_D = 0.4$, the outcome is an equilibrium state with clear-cut between-group polarization. The members of both groups develop a maximally positive attitude about their ingroup and a maximally negative attitude about the outgroup, with no room for individual deviation from these extreme positions and no further change once this state has been reached. Under socially influenced intergroup attitudes, we observe instead what could be called 'mixed instable attitude polarization', with co-existence of extreme and moderate opinions within both groups. Especially, both groups can contain a minority of 'renegades' with negative opinions about their ingroup in these conditions and, moreover, dynamics never settled down to an equilibrium state in the time period simulated in experiment 1[3].

**Experiment 2: intergroup attitudes and large initial disagreement between groups**

Repulsive influence is more likely to occur between individuals with large opinion disagreement. The relatively moderate initial ingroup-bias imposed in experiment 1 may therefore be a critical ingredient for the low degree of between-group polarization that we observed in the condition with socially influenced intergroup attitudes. In experiment 2, I tested whether the differences between fixed xenophobia and flexible intergroup attitudes remain, if in both groups there is a stronger ingroup-bias at the outset.

For this, I used an initial distribution in which the distance between the mean opinions of both groups is about $|\overline{o_{g=0}} - \overline{o_{g=1}}|$ = 0.62, approximately 4 times as large as in experiment 1. All other parameters remain the same than in experiment 1. For comparison, I chose for experiment 2 the parameters of the Beta distributions such that the standard deviations within each of the two groups are the same than in experiment 1 (approximately 0.12) and the distributions of both groups were again symmetrical to the population mean. This resulted in initial distributions[4] drawn from Beta(8.58,2) and Beta(2, 8.58) with mean values of about 0.811 and 0.189 for groups 1 and 0 respectively. Figure 9 visualizes the underlying probability density functions together with the aggregate distribution for the population as a whole.

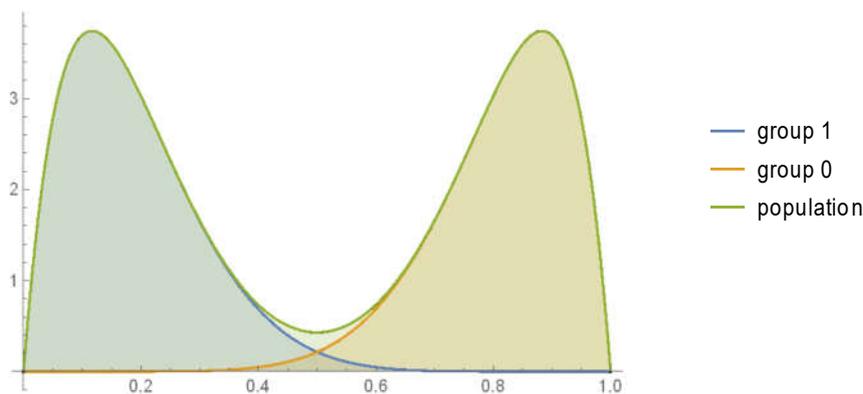

**Figure 9. Initial opinion distribution in experiment 2 (group 1: Beta(8.58,2), group 0: Beta(2, 8.58)).**

Experiment 2 confirmed the expectation that a higher initial disagreement between the group elicits also higher levels of between-group polarization. Comparing the results reported in figure 10 with experiment 1, it is apparent that this held for both fixed xenophobia and socially influenced intergroup attitudes. In the condition with fixed xenophobia, between-group polarization was at its maximum level across the entire spectrum of values for $\beta_D$ that were simulated. With socially influenced intergroup attitudes, between-group polarization only fell discernably below its maximal level when $\beta_A$ exceeded a level of approximately 0.8, with the lowest level obtained at $\beta_A$=1, when the average of $\left|\overline{o_{g=1}} - \overline{o_{g=0}}\right|$ measured after 30.000 simulation events dropped to about 0.65, with an average population polarization *P* of about 0.495.

---

[3] In fact, for the conditions of experiment 1 shown in figure 5 to 7, I found in further simulation runs that dynamics failed to converge within 1 billion (10⁹) simulation events for $\beta_A = 1$, while equilibrium was in some cases reached for $\beta_A = 0.92$ within about 100 million simulation events.
[4] Parameter values for the Beta distributions rounded to two decimals.

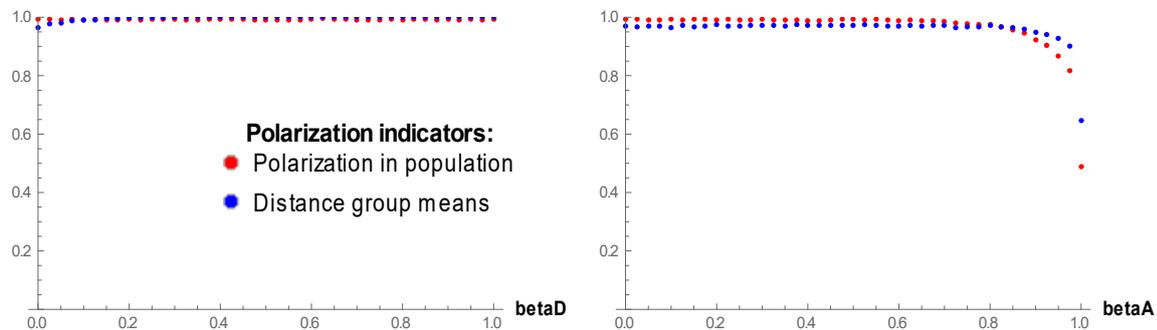

**Figure 10. Effects of relative weight of intergroup attitude on between-group polarization in experiment 2. Polarization indicators averaged across 100 realizations per level of $\beta_D$ and $\beta_A$, respectively. Results after 30.000 simulation events per run. Left: fixed xenophobia condition, Right: socially influenced intergroup attitudes condition.**

Like in experiment 1 we observe that the qualitative effects of for $\beta_D$ and $\beta_A$ on between-group polarization are different. At the same time, the extent becomes much smaller to which socially influenced intergroup attitudes reduce polarization between groups as compared to fixed xenophobia. The initial disagreement between members from different groups is sufficiently high to almost certainly trigger the self-reinforcing dynamics of mutual differentiation between the groups. Only for very high levels of $\beta_A$ this does not always happen, because here neither disagreement nor being demographically dissimilar add much to the discrepancy individuals experience. Still, inspection of the aggregated distribution of opinions over all 100 runs for $\beta_A$ =1 shows that about half of the members of both groups adopted a maximally negative outgroup attitude in experiment 2 after 30.000 simulation events, as compared to less than 5% in experiment 1.

**Experiment 3: renegades and reversed intergroup polarization**

Experiment 2 showed that strong initial ingroup-favoritism in both groups largely undermines a possible effect that renegades could have on preventing between-group polarization. With the initial distributions employed in experiment 2, renegades rarely occur and if they occur their bias towards the outgroup is typically only very small. This does not suffice to let renegades substantially influence the majority of ingroup-lovers within their own group and prevent this majority from 'radicalizing'. In experiment 3, I assess whether renegades can turn the tide if there are more of them from the outset. More specifically, I explore in experiment 3 the dynamics of between-group polarization if the initial opinion distributions show a much larger variance than before, so that the proportion of renegades is higher in both groups.

Figure 11 shows the Beta distributions used in experiment 3. With about 0.067, the initial distance in group means is approximately half as large as in experiment 1, but the standard deviation within both groups is about twice as large (0.229 experiment 3 vs. 0.115 experiments 1 and 2).

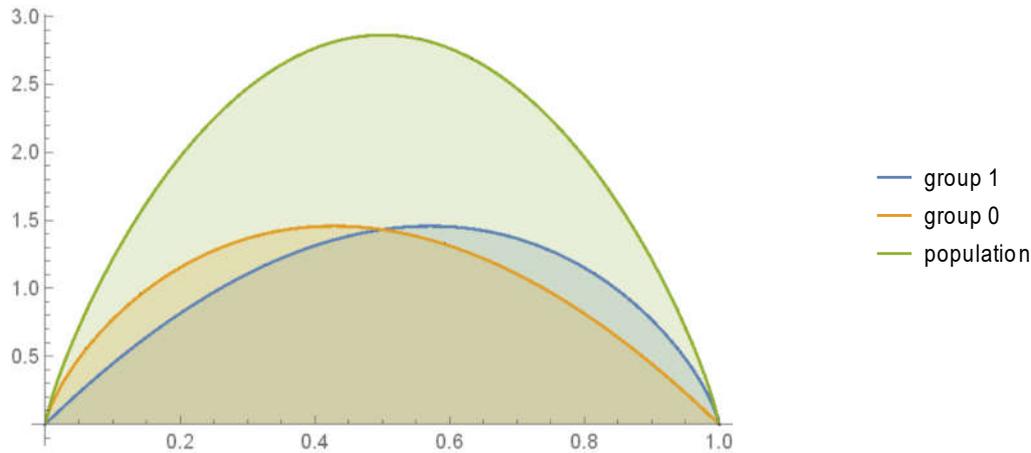

**Figure 11. Initial opinion distribution in experiment 3 (group 1: Beta(2,1.75), group 0: Beta(1.75, 2)).**

In experiment 3, I want to assess whether a larger share of renegades can prevent between-group polarization under conditions for which it is likely to occur otherwise. As a conservative test, socially influenced intergroup attitudes will in experiment 3 be gradually replaced by fixed xenophobia, rendering it increasingly difficult to prevent between-group polarization. That is, starting from a baseline in which social influence is exclusively moderated by intergroup attitudes ($\beta_A$=1), the relative weight of fixed xenophobia will be increased from $\beta_D$= 0 to $\beta_D$= 0.2, and finally $\beta_D$= 0.4. Furthermore, to prevent that the relatively small initial attitude difference between the groups makes consensus too easy to reach, it is assumed throughout experiment 3 that the (low) disagreement in attitudes between members from different groups has no impact on the discrepancy individuals perceive ($\beta_O$= 0, $\beta_D + \beta_A = 1$).

Notice that with this approach it is assumed that individuals can simultaneously experience a fixed degree of outgroup aversion and at the same time be socially influenced to develop a positive attitude towards the outgroup. This captures the idea that although negative views of outgroups can be 'unlearned', there are also simultaneous psychological processes and social-structural reasons (like speaking similar languages or sharing familiar social norms) that let us be more attracted to more similar people, all other things being equal (e.g. (Byrne 1971)). The relative values of $\beta_D$ and $\beta_A$ can be seen as modelling how deeply entrenched in a given social setting are conditions that sustain xenophobia relative to social influence towards more positive intergroup attitudes.

First, I establish a baseline condition with a low share of renegades, to then show how the presence of renegades changes dynamics. For the baseline I use the same initial distribution than for experiment 1. Figure 12 shows opinion dynamics averaged across 100 independent runs for the baseline situation, comparing the scenario $\beta_A$=1 (left) with $\beta_D = 0.2, \beta_A = 0.8$ (right).

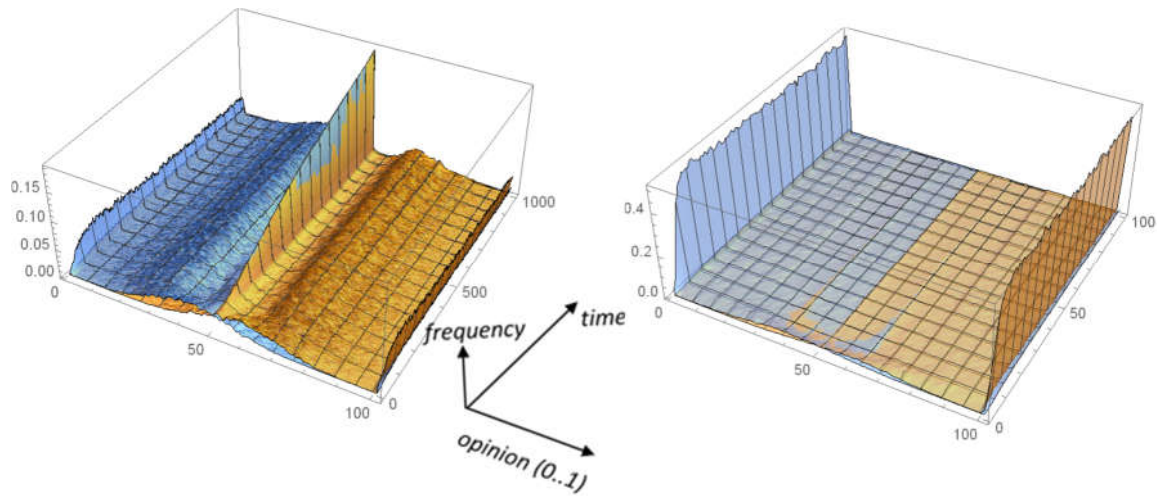

**Figure 12. Baseline condition for experiment 3 (using initial distribution of experiment 1). Change of aggregated opinion distributions over 100 independent realizations for $\beta_A = 1$ (left) and $\beta_A = 0.8$ , $\beta_D = 0.2$ (right) with socially influenced intergroup attitudes ($\beta_O = 0$). Blue: group 0, Orange: group 1.**

The baseline scenario in figure 12 shows how a relatively small increase of the weight of fixed xenophobia from $\beta_D = 0$ to $\beta_D = 0.2$ suffices to push the population from a state in which moderate attitudes dominate into a state with a considerable level of between-group polarization. With $\beta_D = 0.2$, in both groups a share of about 40% of all members moves on average to the attitude of maximal ingroup-love and outgroup-hate, with an average distance between the group means of about 0.69. While this outcome could still be seen as moderate between-group polarization, a further increase of the weight of fixed xenophobia to $\beta_D = 0.4$ pushes the population into a state of nearly maximal between-group polarization (average $|\overline{o_{g=0}} - \overline{o_{g=1}}| \approx 1$).

Next, the baseline condition is compared to the scenario with a larger initial share of renegades (see figure 11). Figure 13 below shows individual trajectories of typical single runs for $\beta_D = 0$, $\beta_D = 0.2$ and $\beta_D = 0.4$, respectively ($\beta_O = 0$, $\beta_D + \beta_A = 1$). Figure 14 depicts the corresponding opinion distribution averaged over 100 runs per condition. The results reveal that the larger share of initial renegades fundamentally changes how fixed xenophobia affects the dynamics of between-group polarization. Arguably, the pattern we observed can be characterized as 'reversed intergroup polarization'. Members of the two groups end up disagreeing like they do with 'regular' between group polarization, but this time both like the outgroup and dislike the ingroup. This can also be seen from comparison of the absolute difference between group means $|\overline{o_{g=0}} - \overline{o_{g=1}}|$, with the signed difference $\overline{o_{g=1}} - \overline{o_{g=0}}$. While the latter is negative for $\beta_D = 0.2$ and $\beta_D = 0.4$, the former is positive and increasing as $\beta_D$ increases from 0 upwards.

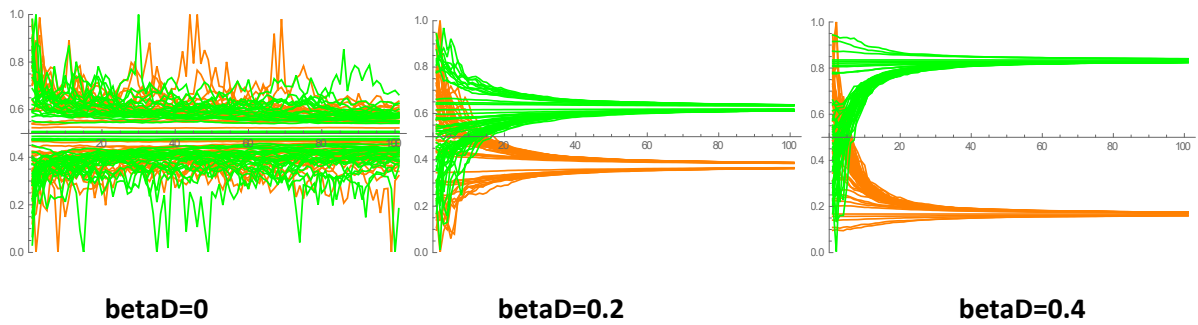

**betaD=0**     **betaD=0.2**     **betaD=0.4**

**Figure 13. Individual trajectories with large share of initial renegades ($\beta_O = 0$, $\beta_D + \beta_A = 1$)**

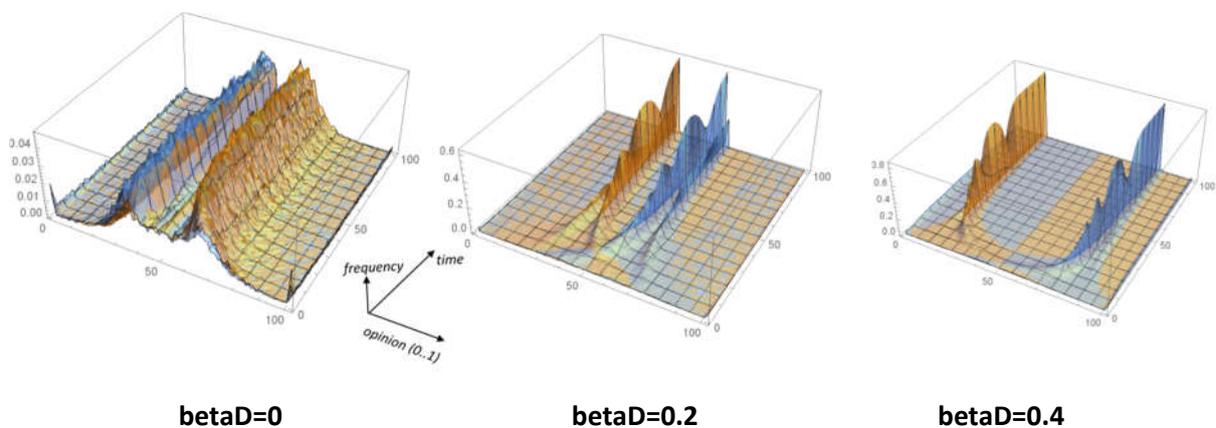

**betaD=0**     **betaD=0.2**     **betaD=0.4**

**Figure 14. Change of opinion distributions averaged over 100 runs under large share of initial renegades ($\beta_O = 0$, $\beta_D + \beta_A = 1$).**

Higher levels of fixed xenophobia increased the extent of ingroup-love and outgroup aversion in both groups when the initial share of renegades was small. Figures 13 and 14 show how this effect is reversed by a high share of initial renegades. Now, from $\beta_D = 0.2$ on, in both groups a consensus develops on 'universal outgroup-love' (o < 0.5 for group 1, o > 0.5 for group 0). Paradoxically, the higher the weight of fixed xenophobia, the more positive the attitude towards the outgroup becomes (and the more negative the attitude towards the ingroup), on which all members of the same group eventually agree. The individual trajectories depicted in figure 13 illustrate how this happens. For $\beta_D = 0.2$ and $\beta_D = 0.4$, initial renegades within both groups change their attitudes much less than initial ingroup-lovers do. The reason is that renegades have a negative attitude towards their own group and are therefore not open for assimilative influence from fellow ingroup members. At the same time, ingroup-lovers wish to assimilate towards the attitudes of their fellow ingroup members, including the renegades. Thus, the influence relation between ingroup-lovers and renegades is asymmetrical. While renegades stay their course, ingroup lovers adopt the negative view of the renegades about their own ingroup, because they wish to agree with their fellow ingroup members. This counter-intuitive dynamic is further strengthened by renegades' attraction towards the attitudes of outgroup members, and ingroup lovers' desire to differentiate from outgroup members. This explains the initial shift of extreme renegades towards slightly less negative views of their ingroup that we can observe for $\beta_D = 0.2$ and $\beta_D = 0.4$ in figure 13. The reason is not assimilation towards the ingroup, but it is assimilation towards the majority of the outgroup members who simultaneously develop an increasingly critical attitude towards their own group.

The position of the equilibrium attitudes to which individuals settle in experiment 3 can be best understood by an analysis of the weight functions *f* imposed by the values for $\beta_O, \beta_D, \beta_A$ that we use in the experiment. Figure 15 charts the weight functions for the three conditions of experiment 3.

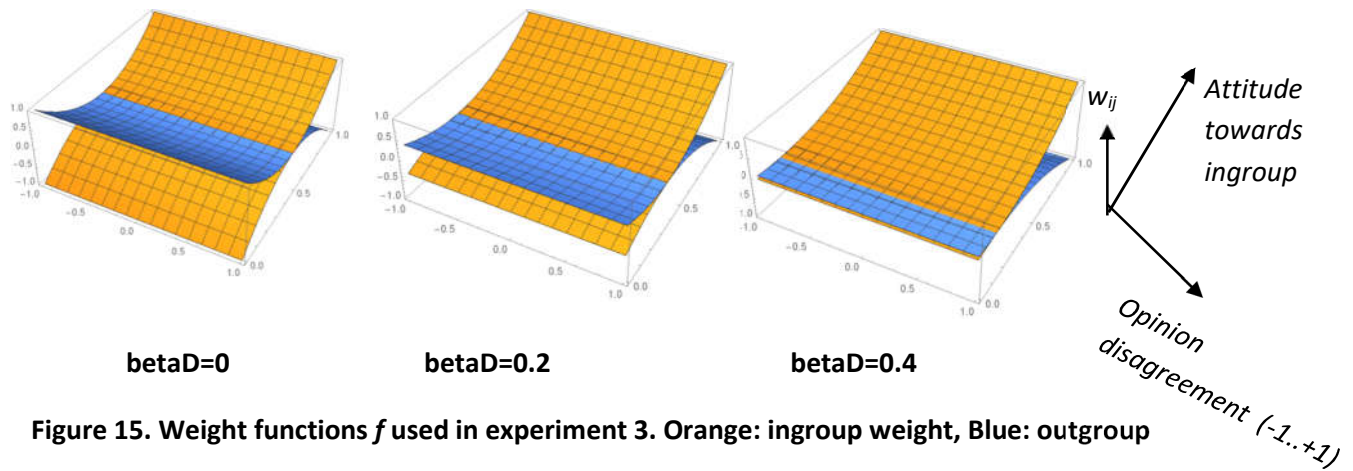

**betaD=0**  **betaD=0.2**  **betaD=0.4**

**Figure 15. Weight functions *f* used in experiment 3. Orange: ingroup weight, Blue: outgroup**

Figure 15 makes clear that in all three conditions there is a value for the intergroup attitude *o* at which the influence weights of both ingroup members and outgroup members are zero. This is the level of *o* at which the two surfaces in the figures intersect. If all individuals within a group adopt this value of *o*, they are no longer susceptible to outside influences and have reached an equilibrium state. Figure 15 shows how increasing $\beta_D$ pushes this equilibrium intergroup attitude towards an increasingly negative opinion about the ingroup. Substantively, this can be interpreted as the interplay of two counterbalancing forces. Fixed xenophobia leads an individual to differentiate from outgroup members. However, if the same individuals hold a favorable socially influenced attitude towards the outgroup, this can neutralize the force towards differentiation and thus result in an equilibrium situation. The stronger fixed xenophobia, the more positive the attitude towards the outgroup needs to be to induce a situation where the net influence of an outgroup member is zero. Conversely, fixed xenophobia makes an individual more open to assimilate towards the opinion of an ingroup member (by helping to reduce discrepancy below the critical level needed for a positive influence weight) but an unfavorable attitude towards the ingroup can neutralize this assimilative force with an equally strong repulsive pressure. The appendix of this paper shows analytically how higher $\beta_D$ shifts the equilibrium intergroup attitude downward in this condition.

The stronger the force of fixed xenophobia, the more favorable the attitude towards the outgroup needs to be to obtain an intergroup attitude in equilibrium. The limiting case is $\beta_D = \beta_A$. When fixed xenophobia and socially influenced intergroup attitude have an equally strong impact on perceived discrepancy, the only stable attitude is a maximally negative view of the ingroup. This holds also when there is an additional impact of disagreement ($\beta_O > 0$) as demonstrated by figure 16 for the case where all three sources of discrepancy have the same impact on the influence weight $w_{ij}$. The figure shows that the influence weight of an ingroup member always exceeds the weight of an outgroup member except for the limiting case of *o*=0 in which both weights are equal for the same level of disagreement. Notice that not both weights are always zero in this case, this only happens if in addition the disagreement between *i* and *j* in the attitude *o* is exactly 0.5.

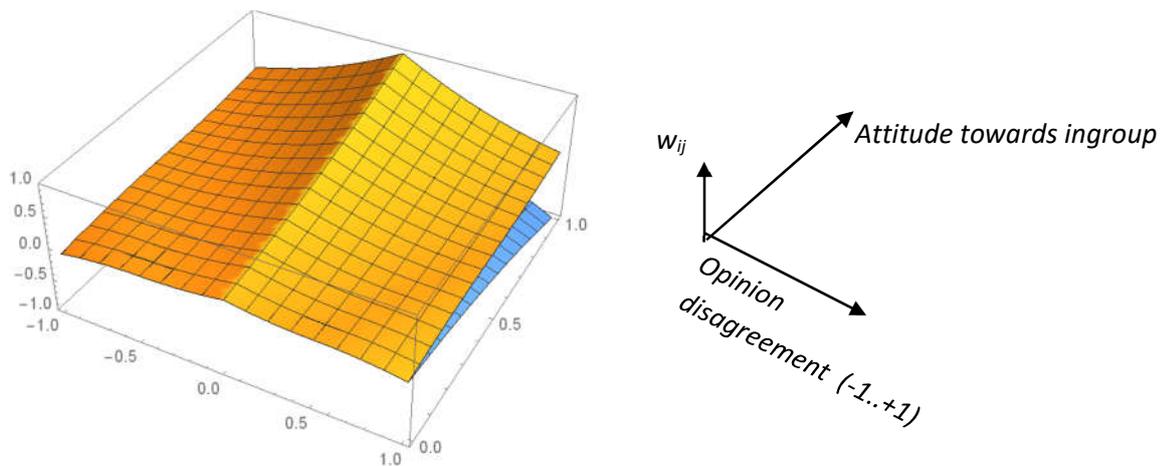

**Figure 16. Weight function *f* for $\beta_D = \beta_A = \beta_O = {}^1/_3$. Orange: ingroup weight, Blue: outgroup weight.**

The situation depicted by Figure 16 suggests that for this parameter vector there can be an equilibrium state with extreme 'reversed intergroup polarization'. This is the state in which both groups adopt a maximally negative view of their own group. In this case, members of both groups are not pulled towards the opposite view of the outgroup they love, because their disagreement with this outgroup is too large. This state is likely to emerge because initial renegades are immune to influence from ingroup members with a favorable view of their own group, while the latter are attracted towards the opinion of the renegades. As a test, I conducted a simulation of this scenario, using the initial distribution of experiment 3. Figure 17 shows two typical runs of this condition. In one of these runs (right graph), a quick reversal of the initial tendencies of both groups occurs such that almost perfect 'reversed intergroup polarization' arises and stabilizes. In the other run (left), the population moves into 'regular' between group polarization. However, some renegades join the other side. As the left part of figure 17 shows, this prevents the dynamics from settling into an equilibrium because ingroup-lovers in both camps are pushed to differentiate from the attitudes of the renegades who simultaneously belong to the outgroup and happen to agree with them.

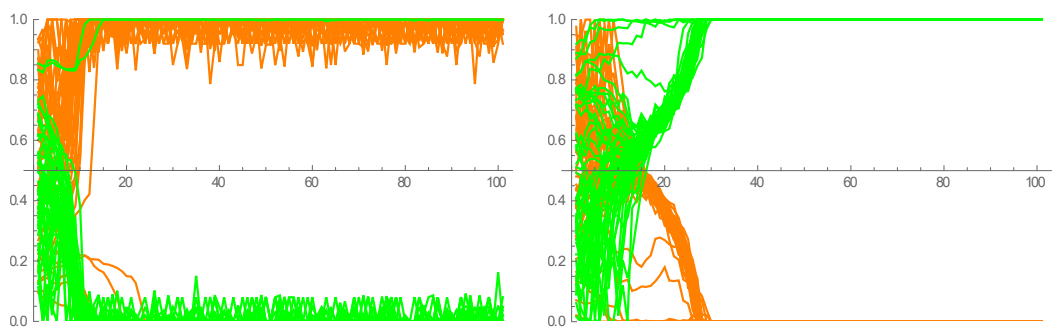

**Figure 17. Two typical simulation runs for $\beta_D = \beta_A = \beta_O = {}^1/_3$. Orange: group 1, Green: group 0.**

Figure 18 shows the corresponding dynamics of the average distribution of opinions aggregated over 100 realizations (left) and the average trajectories of the mean opinions within both groups and the population as a whole.

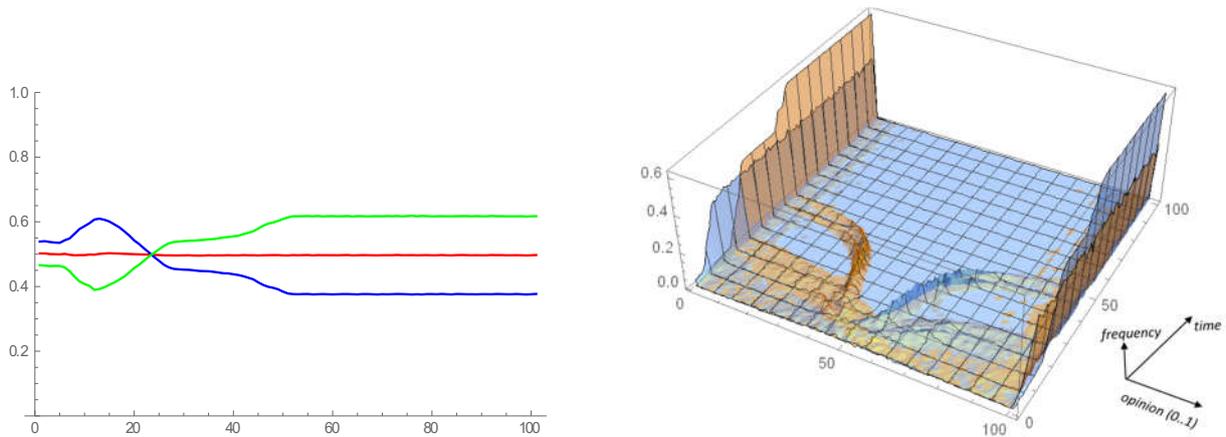

**Figure 18. Average change of group means (left) and opinion distribution (right) based on 100 realizations for $\beta_D = \beta_A = \beta_O = {1}/{3}$ . Left graph: group 0=green, group 1=blue, population=red. Right graph: Orange: group 1, Blue: group 0.**

Figure 18 confirms how a sufficient critical mass of renegades at the outset can change the dynamics of intergroup polarization. On average, more individuals develop into ingroup-haters and outgroup-lovers in this condition than the other way around. Still, on average both groups contain a variety of different attitudes, caused by the instances in which the 'mixed between-group polarization' occurs that is shown in the left part of Figure 17.

## 4. Discussion and conclusion

Formal models of opinion dynamics hitherto consider intergroup attitudes mainly in terms of fixed xenophobia. This implies that individuals differentiate more likely from outgroup members than from ingroup members, and assimilate more likely towards ingroup members than towards outgroup members. These models are prone to generate the outcome of bi-polarization of opinions aligned with group boundaries. However, these models neglect an important insight from research on intergroup relations. Intergroup attitudes are themselves subject to social influence from both ingroup members and outgroup members.

In this paper, a model was proposed that extends previous models of opinion dynamics in demographically diverse groups. The model integrates the feedback loop between social influence on intergroup attitudes and the effect that intergroup attitudes have on influence from ingroup members and outgroup members. Simulation experiments show that the effects of fixed xenophobia on intergroup polarization can profoundly differ from those of socially influenced intergroup attitudes. Broadly, I find that not only are the conditions for intergroup polarization more restrictive when intergroup attitudes are open to social influence. Also, when intergroup attitudes have a stronger impact on social influence this does not necessarily increase between-group polarization, in contrast with models assuming fixed xenophobia. Instead, a model with socially influenced intergroup attitudes can generate richer dynamics and emergent opinion distributions, including outcomes in which both groups contain ingroup-lovers, individuals with a moderate intergroup attitude, and ingroup-haters. Especially, when there is a sufficient initial critical mass of 'renegades' who hold a more favorable attitude of the outgroup than of the ingroup, this critical mass can trigger a dynamic of reversed intergroup polarization in which eventually all members of a group side with the renegades in being critical of their ingroup.

The simulation experiments show that in modelling intergroup opinion dynamics, we can not safely neglect the possibility that intergroup attitudes not only affect social influence but also are subject to social influence. At the same time, these simulation results should not readily be seen as guidance for empirical research on intergroup relations. The model presented here constitutes a first step in integrating social influence with intergroup attitude dynamics and it relies on a number of simplifications and abstractions that need to be carefully explored in future work. In what follows, the most important simplifications are discussed.

First, it is assumed in the current model that there is only one attitude dimension that is subject to social influence, and this attitude dimension simultaneously is the intergroup attitude. This simplification was chosen to focus on what happens to opinion dynamics if intergroup attitudes are socially influenced. At the same time, both earlier modelling work (e.g. Baldassarri and Bearman 2007; Flache and Mäs 2008a,b; Huet and Deffuant 2010; Macy and Flache 2009) as well as empirical evidence (e.g. Gentzkow 2016) suggest that opinion differences on issues which are not necessarily linked to group identities can align over time with 'fixed' differences between groups, but opinions on such issues may also create bridges between groups. Future work should explore whether socially influenced intergroup attitudes mitigate the degree to which demographic differences can elicit between-group polarization also on other issues than intergroup attitudes, and under which conditions this is possible.

A second simplification is the assumption that there is an inverse relation between ingroup and outgroup attitudes. In the model presented here, a more positive attitude towards the ingroup implies a less positive attitude towards the outgroup and vice versa. Drawing on social identity theory (Tajfel and Turner 1979) empirical research on intergroup attitudes found support for such a negative relation between ingroup and outgroup attitudes (e.g. (Mummendey, Klink, and Brown 2001) under certain conditions. However, this model assumption should be seen as simplification of an empirically more complex relationship between in- and outgroup attitudes (see e.g. Brewer 1999; Phinney, Jacoby, and Silva 2007) that can be incorporated in this modelling framework in future research. Changing this assumption may, for example, yield different insights about the role of renegades. If renegades can at the same time have a favorable view of their own group as well as of an outgroup, a critical mass of renegades may less readily lead to the reversed between group polarization that has been shown for the present model, because renegades would be open to influence from both ingroup and outgroup members.

A third avenue for future research is exploration of alternative assumptions about the influence process as such. The model presented here belongs to the broader class of models that combine assimilative and repulsive influence (e.g. Baldassarri and Bearman 2007; Flache and Macy 2011; Jager and Amblard 2005; Macy et al. 2003; Mark 2003). These models are prone to generate bi-polarization under a large range of conditions (e.g. Flache et al, forthcoming). While there is some evidence in experimental and field research for repulsive influence (Hovland, Harvey, and Sherif 1957; Liu and Srivastava 2015), recent experimental research also showed that repulsive influence may be less easily triggered in social interactions than these models suggest (Takács, Flache, and Mäs 2016). Alternative assumptions about influence processes studied in the literature include "bounded confidence" (Deffuant et al. 2000; Hegselmann and Krause 2002) influence constrained by homophily (Axelrod 1997), influence modified by "biased assimilation" (Dandekar, Goel, and Lee 2005; Mäs et al. 2013), or assimilative influence modified by a persistent pull towards fixed initial tendencies (Friedkin 2001) or by a "strive for uniqueness" (Mäs, Flache, and Helbing 2010). Broadly, most of these alternative specifications tend to generate opinion diversity more in the form of clustering around a broader range of alternatives in the opinion space, rather than bi-polarization. Future work should thus analyze whether and how using alternative specifications of the influence process affects the fundamental differences between fixed xenophobia and socially influenced intergroup attitudes that we identified for the present model.

Finally, the model presented here specifies group identities as fixed individual properties. This portrays a world in which the group identity of an individual is always observable and can never change. Moreover, it includes the assumption that all individuals agree about who belongs to which group. However, social identity theory and research on the formation of social and ethnic boundaries (Bourdieu 1984; Wimmer 2008) shows how individuals themselves can change their identification with social categories and, moreover, outsiders can change their view about the group to which someone belongs and of the salience of social distinctions in society. Future extensions should consider the possibility that individuals modify their identification with and perception of social categories. This would, for example, allow to include the possibility that renegades do no longer see themselves as members of their ingroup or are not perceived as 'proper' ingroup members any more by others. This may diminish the influence renegades can have on ingroup members.

The model presented in this paper employs a number of simplifications that suggest directions for future research. This notwithstanding, it points to an insight hitherto overlooked by models which construe intergroup attitudes as fixed xenophobia. Group identities that differentiate between ingroup and outgroup members do not necessarily foster opinion polarization between groups. When intergroup attitudes are themselves subject to social influence, they may prevent polarization between groups if the extent of initial ingroup-bias is moderate. Or, they may lead to a complex landscape of emergent attitudes about in- and outgroups, including extreme ingroup-love, moderate intergroup attitudes and ingroup-hate within all groups at the same time. This suggests that the model proposed here opens up a wide range of possibilities for future theoretical investigation of the role of socially influenced intergroup attitudes in opinion dynamics.

**Appendix**

In this appendix, some analytical results are derived for the conditions under which an intergroup attitude *o* can be stable. These results will be related to outcomes of the simulation experiments presented in the main paper.

To begin with, conditions are derived for the situation where only fixed xenophobia and socially influenced intergroup attitudes have an effect on perceived discrepancy, i.e. $\beta_O = 0$.

**The case of $\beta_O = 0$**

Under which conditions does an intergroup attitude $o_{it}$ constitute an equilibrium attitude in the sense that the influence of both ingroup-members and outgroup-members on individual *i* is zero?

Given the monotonicity of the weight function (equation [4]), for a given intergroup attitude $o_{it}$, the weight $w_{ijt}$ towards an individual *j* is zero if and only if *i* experiences a discrepancy of $d_{ijt} = \frac{1}{2}$ with *j*.

Thus the condition under which $o_{it}$ is an equilibrium attitude is equivalent to the condition that the discrepancy *i* experiences are exactly 0.5 both for an ingroup member *j* as well as for an outgroup member *j*. Notice that given $\beta_O = 0$, the attitude held by *j* is irrelevant for the discrepancy. Equations [A1] and [A2] give the discrepancies that follow from equation [4] for an ingroup member [A1] and an outgroup member [A2], respectively, given $\beta_O = 0$. The time index *t* is omitted, because this concerns equilibrium conditions. Furthermore, it is assumed without loss of generality that $o_i$ is the attitude about the outgroup of *i*.

$$d_{in} = \beta_A o_i \qquad [A1]$$

$$d_{out} = \beta_D + \beta_A(1 - o_i) \qquad [A2]$$

Suppose $d_{in} = \frac{1}{2}$. From [A1] it follows that this is equivalent with the intergroup attitude taking the value of $o_i^{in}$ given by equation [A3]. Correspondingly, it follows from [A2] that $d_{out} = \frac{1}{2}$ is equivalent with an intergroup attitude $o_i^{out}$ given by [A4].

$$o_i^{in} = \frac{1}{2\beta_A} \qquad [A3]$$

$$o_i^{out} = 1 + \frac{2\beta_D - 1}{2\beta_A} \qquad [A4]$$

Finally, we can derive from [A3] and [A4] a condition for which $o_i^{in} = o_i^{out}$. It follows that this is true if and only if $\beta_A = 1 - \beta_D$, which was the condition imposed from the outset by $\beta_O + \beta_D + \beta_A = 1$ and $\beta_O = 0$. Thus, for $\beta_O = 0$, both ingroup weight and outgroup weight are zero for the same equilibrium attitude $o_i^* = o_i^{in} = o_i^{out}$.

Equations [A3] and [A4] provide a proof that explains one of the results from experiment 1 reported in the main paper. If $\beta_A = 1$, then the equations yield $o_i^{in} = o_i^{out} = \frac{1}{2}$, which shows that $o_{it} = \frac{1}{2}$ is always an equilibrium attitude for *i* if $\beta_A = 1$. This explains why in experiment 1 we see over time an increasing number of individuals adopt $o_{it} = \frac{1}{2}$ if $\beta_A = 1$ (see e.g. figures 5, 6, 7 and also figure 12 for experiment 3).

Equations [A3] and [A4] also explain the reversed intergroup polarization found in Experiment 3. More specifically, the equations explain why, given $\beta_O = 0$, the intergroup attitudes towards which both groups move constitute increasing ingroup-hate and outgroup-love as $\beta_D$ shifts from 0 to 0.2 and 0.4 in Experiment 3 (Figures 13 and 14 in the main paper). For illustration, figure A1 shows how $\beta_A$ affects the equilibrium outgroup-attitude $o_i^*$ that follows from both [A3] and [A4].

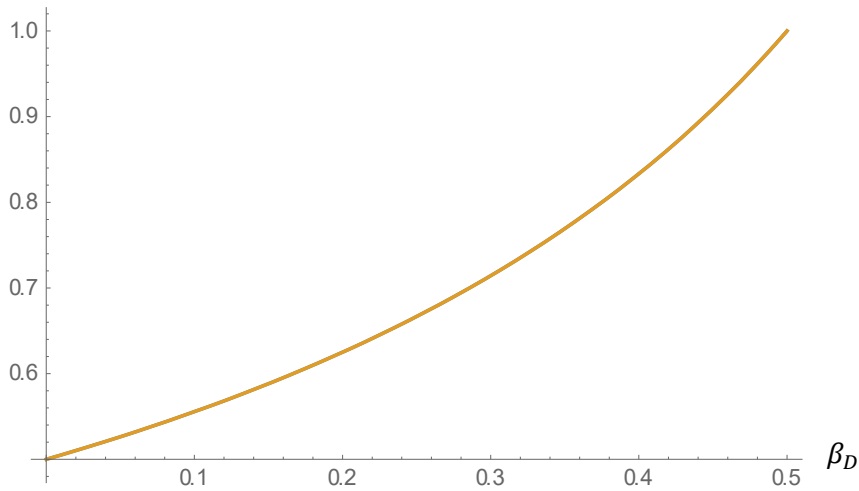

**Figure A1. Equilibrium outgroup attitude $o_i^*$ as function of $\beta_D$ given by both [A3] and [A4].**

Figure A1 shows that for both groups, the larger $\beta_D$, the farther the equilibrium outgroup attitude shifts away from $o_i^* = 0.5$ towards $o_i^* = 1$. For values of $\beta_D > 0.5$, no feasible solutions exist for [A3] and [A4], because both equations yield attitudes larger than one.

**The case of $\beta_O > 0$.**

Conditions are analyzed for equilibria in which there is perfect consensus within both of the two groups, but not necessarily between them. For this, it is useful to distinguish between the attitudes in which members of group 0 and group 1 respectively experience discrepancy $d = \frac{1}{2}$ towards both ingroup members and outgroup members. From now on, it is assumed that the attitude *o* expresses the attitude towards group 1.

Let $d_{in}^0$ denote the discrepancy that members of group 0 experience when interacting with an ingroup member. Given that ingroup members fully agree with the focal individual *i*, equation [3] implies the value of $d_{in}^0$ as given by [A5].

$$d_{in}^0 = \beta_A o_{in} \qquad [A5]$$

Solving [A5] for the condition of zero weight or, $d = \frac{1}{2}$, yields the condition for the equilibrium attitude $o_{in}^0$ of group 0 given by [A6].

$$o_{in}^0 = \frac{1}{2\beta_A} \qquad [A6]$$

Notice that this condition is the same than for $\beta_O = 0$. This follows from the premise that all group 0 members have the same attitude towards group 1, such that their mutual disagreement is zero.

The discrepancy that members of group 1 experience when interacting with an ingroup member can be similarly obtained from equation [3], as given by [A7].

$$d_{in}^1 = \beta_A(1 - o_i) \qquad [A7]$$

Solving [A7] for $d = \frac{1}{2}$ yields the equilibrium attitude $o_{in}^1$ of group 1 given by [A8].

$$o_{in}^1 = 1 - \frac{1}{2\beta_A} \qquad [A8]$$

The conditions given by [A6] and [A8] are necessary conditions for an equilibrium consensus within both groups, but they may not be sufficient. Possible solutions are further constrained by the requirement that in both groups also the discrepancy towards outgroup members must satisfy $d = \frac{1}{2}$. Equations [A9] and [A10] express the discrepancies towards outgroup members as function of the intergroup attitude $o_i$ and the disagreement $\Delta o$ towards the attitude of the outgroup member for group 0 and group 1, respectively.

$$d_{out}^0 = \beta_O \Delta o + \beta_D + \beta_A(1 - o_i) \qquad [A9]$$

$$d_{out}^1 = \beta_O \Delta o + \beta_D + \beta_A o_i \qquad [A10]$$

Equilibrium requires that $o_i = o_{in}^0$ in [A9] and $o_i = o_{in}^1$ in [A10]. Substituting the term from [A6] for $o_i$ in [A9] and from [A8] for $o_i$ in [A10], and then solving both [A9] and [A10] for the disagreement with the outgroup $\Delta o^*$ that satisfies the equilibrium conditions, we obtain for both equations the same solution:

$$\Delta o^* = \frac{1 - \beta_A - \beta_E}{\beta_O} \qquad [A11]$$

By virtue of the premise $\beta_O + \beta_D + \beta_A = 1$, we thus obtain $\Delta o^* = 1$ as the only feasible level of disagreement between the groups that satisfies these equilibrium conditions. From [A6] and [A8] it follows that this level can in equilibrium only be obtained if $\beta_A = 0.5$, which yields $o_{in}^0 = 1$ and $o_{in}^1 = 0$. Thus with $\beta_A = 0.5$, maximal "reversed intergroup polarization" constitutes an equilibrium.

To be sure, there also other conditions possible under which no change occurs in the population. In particular, the basic influence equations [1] and [2] from the main paper imply that an individual will never change her opinion after interaction with another individual who holds the same attitude. This is, because $\Delta o = 0$ if $(o_{jt} - o_{it}) = 0$, regardless of the influence weight $w$. Thus, every state in which all members of the population have the same attitude is an equilibrium. Furthermore, every state in which there is consensus within each of the two groups can be an equilibrium if the condition is satisfied that for both groups the discrepancy experienced in interaction with outgroup members meets $d = \frac{1}{2}$. Thus, equilibrium also is reached if there is consensus within each of the two groups and in addition:

$$d_{out}^0 = \beta_O \Delta o + \beta_D + \beta_A(1 - o^0) = \frac{1}{2} \qquad [A12]$$

and

$$d^1_{out} = \beta_O \Delta o + \beta_D + \beta_A o^1 = \frac{1}{2} \quad [A13]$$

From the requirement $d^0_{out} = d^1_{out}$ it follows that

$$\beta_A(1 - o^0) = \beta_A o^1$$

Which is equivalent with

$$(1 - o^0) = o^1 \quad [A14]$$

This condition is not sufficient to guarantee equilibrium, but it shows that if there is an equilibrium with perfect consensus within each of the two groups, the attitudes which the two groups adopt are symmetrical to the midpoint of the opinion scale. Figure 13 in the main paper shows some examples of cases where $\beta_O = 0$. Figure A2 below shows two examples of conditions with $\beta_O > 0$. In both runs, the initial distribution was drawn from the Beta distributions used in experiment 3.

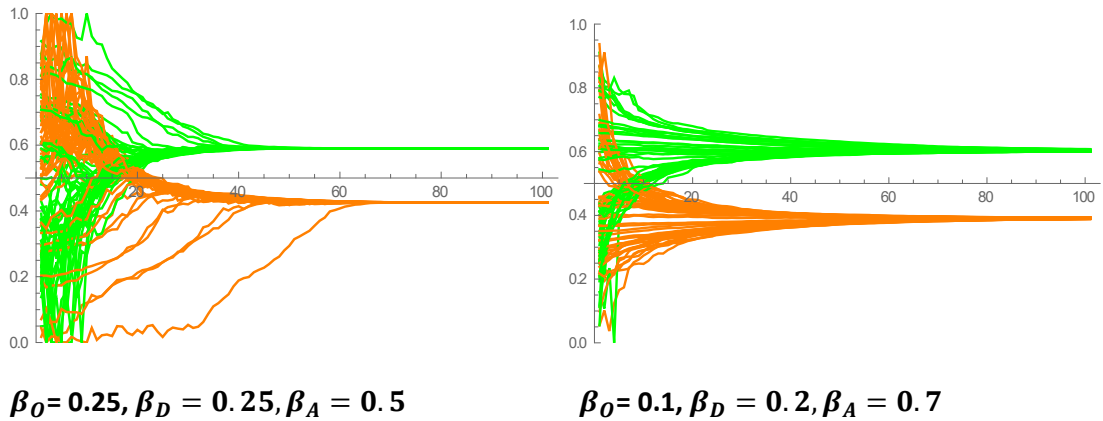

$\boldsymbol{\beta_O}$= 0.25, $\boldsymbol{\beta_D = 0.25, \beta_A = 0.5}$     $\boldsymbol{\beta_O}$= 0.1, $\boldsymbol{\beta_D = 0.2, \beta_A = 0.7}$

**Figure A2. Two examples of runs for conditions with $\boldsymbol{\beta_O} > 0$ generating long-run stable outcomes with consensus within each of the two groups adopting opinions satisfying equation [A14]. Both cases show "mild reversed intergroup polarization". Individual trajectories show emergence of the two group opinions within the first 30.000 simulation events. Both were still stable after running 300.000 simulation events.**

It also possible to give a more precise characterization of the conditions under which equilibria with consensus within each of the two groups can obtain.

To see this, rewrite the conditions given by [A12] and [A13] by expanding the term $\Delta o$. There are two possibilities. In "reversed" intergroup polarization, the equilibrium attitude of group 0 exceeds the one of group 1, i.e. $o^0 \geq o^1$, in "regular" intergroup polarization it is the other way around, i.e. $o^1 \geq o^0$.

First, rewrite [A12] and [A13] for reversed intergroup polarization:

$$d_{out}^0 = \beta_O(o^0 - o^1) + \beta_D + \beta_A(1 - o^0) = \tfrac{1}{2} \qquad \text{[A15]}$$

$$d_{out}^1 = \beta_O(o^0 - o^1) + \beta_D + \beta_A o^1 = \tfrac{1}{2} \qquad \text{[A16]}$$

Solving these two conditions simultaneously for *o⁰* and *o¹* yields:

$$o^0 = \frac{2\beta_A + 2\beta_D - 2\beta_O - 1}{2(\beta_A - 2\beta_O)} \qquad \text{[A17]}$$

$$o^1 = \frac{1 - 2\beta_D - 2\beta_O}{2(\beta_A - 2\beta_O)} \qquad \text{[A18]}$$

However, there is no guarantee that this solution meets the criterion $o^0 \geq o^1$. To assure this, we can simplify $\frac{2\beta_A + 2\beta_D - 2\beta_O - 1}{2(\beta_A - 2\beta_O)} \geq \frac{1 - 2\beta_D - 2\beta_O}{2(\beta_A - 2\beta_O)}$, which yields, after some further rearrangement utilizing that $\beta_A = 1 - \beta_O - \beta_D$, the condition

$$(\beta_O < \tfrac{1}{3}(1 - \beta_D) \,\&\, (\beta_O \leq \beta_D)) \text{ or } (\beta_O > \tfrac{1}{3}(1 - \beta_D) \,\&\, (\beta_O \geq \beta_D)) \qquad \text{[A19]}$$

If condition [A19] is not met, it is still possible that there an equilibrium with regular intergroup polarization arises. This imposes the precondition that $o^1 \geq o^0$, such that finding the equilibrium conditions requires solving the system given by [A20] and [A21] for $o^1$ and $o^0$.

$$d_{out}^0 = \beta_O(o^1 - o^0) + \beta_D + \beta_A(1 - o^0) = \tfrac{1}{2} \qquad \text{[A20]}$$

$$d_{out}^1 = \beta_O(o^1 - o^0) + \beta_D + \beta_A o^1 = \tfrac{1}{2} \qquad \text{[A21]}$$

This yields the following solutions:

$$o^0 = \frac{2\beta_A + 2\beta_D + 2\beta_O - 1}{2(\beta_A - 2\beta_O)} \qquad \text{[A22]}$$

$$o^1 = \frac{1 - 2\beta_D + 2\beta_O}{2(\beta_A - 2\beta_O)} \qquad \text{[A23]}$$

The condition under which [A22] and [A23] satisfy $o^1 \geq o^0$ reduces to $\beta_O \geq \beta_D$.

This shows that for some parameter configurations only reversed intergroup polarization is possible, for others only regular intergroup polarization, and for some both can obtain in equilibrium. Finally, there are also configuration in which both forms of polarization can constitute an equilibrium outcome.

Here are some examples. In all cases, simulations confirmed that the analytically derived equilibrium outcomes indeed constitute stable states of the dynamics.

Example for *regular intergroup polarization*: $\beta_O = 0.2, \beta_D = 0.1, \beta_A = 0.7$.

This meets the condition for $o^1 \geq o^0$ and does not meet the condition for $o^0 > o^1$. The solution obtained is $o^0 = 0.455$ and o1 = 0.545 (rounded to 3 decimals).

Example for *reversed intergroup polarization*: $\beta_O = 0.1, \beta_D = 0.3, \beta_A = 0.6$.

Now the solution is reversed polarization: $o^0$ = 0.75, $o^1$=0.25. Indeed $\beta_O \geq \beta_D$ is not met, thus the condition for $o^1 \geq o^0$ not met. Also $(\beta_O < \frac{1}{3}(1-\beta_D))$ & $(\beta_O \leq \beta_D)$, thus the condition for reversed intergroup polarization is met.

Finally, here an example for which *both types of equilibrium outcomes are possible*: $\beta_O$= 1/2, $\beta_D = 1/3, \beta_A = 1/6$. In this case, reversed intergroup polarization is sustained for $o^0$ =3/5 and $o^1$=2/5. Regular intergroup polarization is sustained for $o^0$=3/7 and $o^1$=4/7. However, this is no guarantee that any of these equilibria is reached from a random start.

Here an example for this parameter configuration in which neither perfect reversed intergroup polarization obtains, nor perfect regular intergroup polarization. This time *N*=10, the initial distribution is drawn from the same Beta distributions that were used in experiment 3.

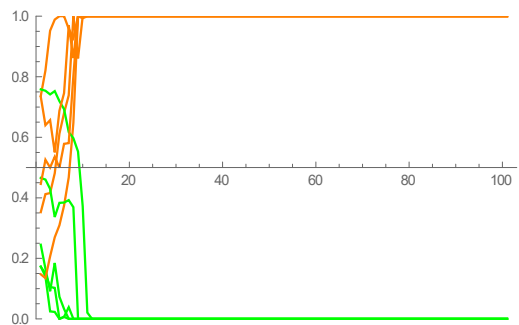

**Figure A3. Extreme regular intergroup polarization for $\beta_O$= 1/2, $\beta_D = 1/3, \beta_A = 1/6$. N=10. 3000 iterations.**